\documentclass[aps,prstab,twocolumn,superscriptaddress,
showpacs,amsmath,amssymb,floatfix,letterpaper,bibnotes]{revtex4} 

\usepackage{graphicx}

\newcommand{\lae}{\lower 2pt \hbox{$\, \buildrel {\scriptstyle <}\over {\scriptstyle\sim}\,$}}
\newcommand{\lbeps}{\mbox{\Large \boldmath $\varepsilon$}}

\begin{document} 

\title{Wide spin resonance with an rf-bunched proton beam} 

\author{V.S.~Morozov}
\altaffiliation[also at ]
{ODU RF, P.O. Box 6369, Norfolk, VA 23508.}
\author{A.W.~Chao} 
\altaffiliation[also ]
{SLAC, 2575 Sand Hill Rd., Menlo Park, CA 94025.}
\author{A.D.~Krisch}
\author{M.A.~Leonova}
\author{J.~Liu}
\author{R.S.~Raymond}
\author{D.W.~Sivers}
\author{V.K.~Wong}
\affiliation{Spin Physics Center, University of Michigan, 
Ann Arbor, MI 48109-1040, USA}

\author{A.M.~Kondratenko}
\affiliation{GOO Zaryad, Russkaya St. 41, Novosibirsk, 630058 Russia}

\date{Submitted to PRL 22oct09; submitted to PRST-AB 22dec09} 

\begin{abstract} 
We recently used an rf solenoid to study the widths of rf spin resonances 
with both \textbf{\emph{unbunched}} and \textbf{\emph{bunched}} beams of 
2.1~GeV/$c$ polarized protons stored in the COSY synchrotron. A map, with  
\textbf{\emph{unbunched}} beam at different fixed rf-solenoid frequencies, 
showed a very shallow possible depolarization dip at the resonance. 
Next we made frequency sweeps of 400~Hz, centered at similar frequencies, which greatly enhanced the dip. 
But, with a \textbf{\emph{bunched}} proton beam, both the fixed-frequency and 
frequency-sweep techniques produced similar maps, and both \textbf{\emph{bunched}} 
maps showed full beam depolarization over a wide region. 
Moreover, both were more than twice as wide as the \textbf{\emph{unbunched}} dip. 
This \textbf{\emph{widening}} of the \textbf{\emph{proton}} resonance due to bunching 
is exactly opposite to the recently observed \textbf{\emph{narrowing}} of \textbf{\emph{deuteron}} resonances 
due to bunching.
\end{abstract} 

\maketitle 

The ability to preserve and precisely control a beam's polarization 
during acceleration and storage is needed to study the spin dependence 
of nuclear and particle interactions~\cite{dgcrabb,spin00,spin02,spin04,spin06}. 
Rf magnets can induce rf spin resonances in storage rings, 
which allow one to manipulate the beam's polarization; 
they also allow detailed spin resonance studies and beam diagnostics. 
Running an rf magnet at different fixed frequencies 
or making small-range sweeps of its frequency 
near a spin resonance can produce a resonance map. 
Such maps can precisely determine the resonance's properties, 
such as its strength, width, central frequency, 
and frequency spread, as well as the beam's 
properties, such as its energy and its momentum spread. 
It was earlier discussed that bunching beams of muons~\cite{orlov,bailey}, 
electrons and positrons~\cite{bukin,serednyakov,derbenev}, or deuterons~\cite{narrowres} 
could narrow a resonance's width and thus increase the measurements' precision. 
We recently used 2.1~GeV/$c$ polarized protons stored 
in the COSY synchrotron for a detailed experimental study of 
both \textbf{\emph{unbunched}} and \textbf{\emph{bunched}} proton beam 
resonances.  

In flat circular rings, each beam particle's spin 
normally precesses around the vertical fields of
the ring's dipole magnets. 
The spin tune $\nu_s = G\:\gamma$ is the number of spin 
precessions during one turn around the ring; 
$G = (g - 2)/2$ is the particle's gyromagnetic anomaly and $\gamma$ is 
its Lorentz energy factor. 
A horizontal magnetic field can perturb the particle's 
stable vertical polarization creating a spin 
resonance~\cite{stora,courant,montague}. 
Rf magnets can induce rf spin resonances~\cite{morozov03,yonehara,
leonova04,morozov05,leonova06,krisch07,leonovaArXiv09,leonovaSPIN08}. 
A proton's rf-induced spin resonance's frequency is
\begin{equation} 
f_r = f_c (k \pm G_p\:\gamma), 
\label{eq:fr} 
\end{equation} 
where $f_c$ is the proton's circulation frequency, $k$ is an 
integer, and $G_p = 1.792\:847$.

The apparatus for this experiment, including the COSY storage 
ring~\cite{maier,lehrach99,stockhorst,lehrach02}, the EDDA 
detector~\cite{schwarz,altmeier}, the electron Cooler~\cite{stein}, 
the low energy polarimeter (LEP)~\cite{chiladze}, the injector cyclotron, 
and the polarized ion source~\cite{eversheim,weidmann,felden} 
were shown in Fig.~4 of ref~\cite{morozov07}. The beam from the polarized 
$H^-$ ion source was accelerated by the cyclotron to 45~MeV 
and then strip-injected into COSY. Before this injection, the LEP measured 
the $H^-$ beam's polarization to monitor its stability. 
 
The 24.5~keV electron Cooler reduced the beam's momentum spread $\Delta p/p$ 
by cooling it, both longitudinally and transversely, for 15~s at the protons' 
45~MeV injection energy. The protons were then accelerated to 
2.1~GeV/$c$, where the rf acceleration cavity was either \textbf{\emph{off}} 
during COSY's flat-top giving an \textbf{\emph{unbunched}} beam, or 
\textbf{\emph{on}} giving a \textbf{\emph{bunched}} beam. 

A spin resonance was induced using an rf solenoid magnet~\cite{morozov08}; 
it was a 25-turn air-core water-cooled copper coil, 
of length 57.5~cm and average diameter 21~cm. Its inductance 
was $41 \pm 3$~$\mu$H. It was part of an RLC resonant circuit, 
which operated near 902.6~kHz, typically at an rf voltage of 5.7~kV~rms. 
The longitudinal rf magnetic field at its center was about 1.17~mT~rms, 
giving an rf $\int\!\! Bdl$ of $0.67~\pm~0.03$~T$\cdot$mm rms.

The cylindrical EDDA polarimeter~\cite{schwarz,altmeier} then measured 
the beam's polarization in COSY. We reduced its 
systematic errors by cycling the polarized source between
the up and down vertical polarization states. 
The measured flattop polarization,
before spin manipulation, was typically about 75\%.

In the COSY ring, the protons' average circulation frequency $f_c$ 
was $1.491\:85$~MHz at 2.1~GeV/$c$, where their 
Lorentz energy factor was $\gamma = 2.4514$. For these parameters, 
the spin tune $\nu_s = G\gamma$ was $4.395$. Thus, Eq.~(\ref{eq:fr}) 
gave that the $k=5$ spin resonance's central frequency should be 
near 
\begin{equation}
f_r = (5-G\gamma)f_c = 902.6~\mbox{kHz}. 
\end{equation}

\begin{figure}[t]
\includegraphics[width=\columnwidth]{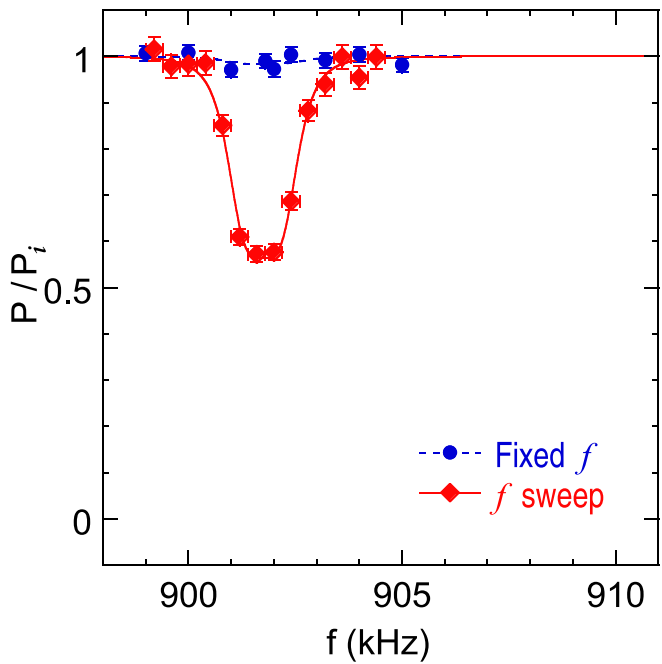}
\caption{
Polarization ratio $P/P_i$, measured at 2.1~GeV/$c$ with 
an \textbf{\emph{unbunched}} proton beam, plotted vs 
the rf solenoid's central frequency $f$, where 
$P$ and $P_i$ are the final and initial 
vertical polarizations, respectively. 
The spin resonance strength $\lbeps$ at full rf-solenoid voltage 
was $\mbox{$31.3 \times 10^{-6}$}$. 
The rf solenoid's ramp-up and ramp-down times $t_R$ were 200~ms; 
its on-time at full $\lbeps$ was $t_{ON}=2$~s. 
The frequency sweep data's range was 400~Hz; its sweep time was 2~s. 
COSY's proton momentum spread is usually less than $10^{-3}$. 
The curves are fits to empirical 2nd-order Lorentzians.  
The errors are purely statistical.
} 
\label{fig:fig1}
\end{figure} 

\begin{figure}[t]
\includegraphics[width=\columnwidth]{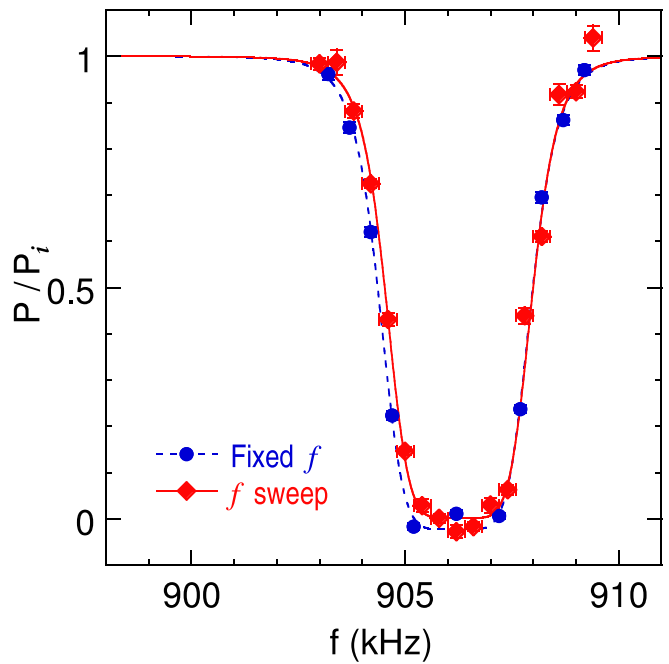}
\caption{Vertical polarization ratio $P/P_i$, 
measured at 2.1~GeV/$c$ with a \textbf{\emph{bunched}} proton beam, 
plotted vs the rf solenoid's central frequency $f$. 
The rf-solenoid parameters are in Fig.~\ref{fig:fig1} caption; 
the beam's synchrotron frequency $f_s$ was 56~Hz.
The curves are fits to empirical 3rd-order Lorentzians.
} 
\label{fig:fig2}
\end{figure} 

We first obtained the rf-induced spin resonance's strength $\lbeps$ 
experimentally~\cite{leonova06,krisch07,leonovaArXiv09,leonovaSPIN08}.  
The polarization was measured after ramping the rf solenoid's frequency 
through the resonance with various ramp times $\Delta t$, while the ramp's 
frequency range $\Delta f$ and voltage were both fixed. 
We then fit these data to the modified~\cite{modifFS} Froissart-Stora 
equation~\cite{stora} with $\lbeps$ as a fit parameter to obtain 
$\lbeps = (31.3~\pm~0.1) \times 10^{-6}$.

A resonance map was then obtained with the beam \textbf{\emph{unbunched}}. 
For different fixed rf-solenoid frequencies $f$ near 902.6~kHz, we linearly 
ramped the rf solenoid's strength from zero to full $\lbeps$ 
during $t_R = 200$~ms; we then held $\lbeps$ fixed during $t_{ON} = 2$~s; 
next we linearly ramped it to zero during $t_R = 200$~ms. 
The resulting measured polarization ratios $P/P_i$ are plotted 
in Fig.~\ref{fig:fig1}. Note that any possible depolarization dip is 
very shallow and the measured final polarization is almost 
consistent with its initial value. This is probably due to 
the proton beam's large momentum spread $\Delta p/p$~\cite{DPnote}. 
We fit these fixed-frequency \textbf{\emph{unbunched}} $P/P_i$ data 
to an empirical 2nd-order Lorentzian function obtaining 
$\chi^2/(N-3)$ of 0.9. 
The fit gave a central resonance frequency $f_r$ of 
$\mbox{$901.8 \pm 0.8$ kHz}$ and a width $w$ of 
$\mbox{$2 \pm 2$ kHz}$~FWHM. The large uncertainties in 
this fixed-frequency fit are due to the possible dip being very shallow. 

To enhance the depth of the \textbf{\emph{unbunched}} resonance and 
thus improve the precision of its frequency and width measurements, 
we made 400~Hz rf-solenoid frequency sweeps near the resonance. 
We again ramped the rf solenoid's strength from 
zero to full $\lbeps$ in 200~ms while holding its 
frequency fixed; we then ramped its frequency by 
400~Hz in 2~s; we next ramped $\lbeps$ down 
to zero while again holding the frequency fixed. 
We made many such sweeps adjacent to one another 
to cover the entire resonance range. 
The resulting data are plotted against the sweeps' 
central frequencies in Fig.~\ref{fig:fig1} along with 
the fixed-frequency data. Each point's frequency-sweep 
range is indicated by a horizontal bar. 

As seen from Fig.~\ref{fig:fig1}, the frequency-sweep 
technique greatly enhanced the \textbf{\emph{unbunched}} 
map's depolarization dip allowing a precise 
determination of the resonance's location and width. 
Fitting the frequency-sweep data in Fig.~\ref{fig:fig1} 
to a 2nd-order Lorentzian yielded 
$f_r$ of $\mbox{$901.7 \pm 0.2$ kHz}$ and $w$ of 
$\mbox{$1.7 \pm 0.2$ kHz}$~FWHM with $\chi^2/(N-3)$ of 0.8. 
Note that the errors in $f_r$ and $w$ are dominated by 
the frequency-sweep's $\pm \, 200$~Hz size. 
Averaging the fixed-frequency and frequency-sweep results 
gave $\mbox{$f_r = 901.7 \pm 0.2$ kHz}$ 
and $\mbox{$w = 1.7 \pm 0.2$ kHz}$~FWHM for \textbf{\emph{unbunched}} 
beam. 

We next used the procedure described above to obtain 
fixed-frequency and frequency-sweep maps of a 
\textbf{\emph{bunched}} proton beam with a 
synchrotron frequency $f_s$ of 56~Hz. These data are 
shown in Fig.~\ref{fig:fig2} using the same scale and 
notation as in Fig.~\ref{fig:fig1}. 
We fit both Fig.~\ref{fig:fig2} resonance maps 
to empirical 3rd-order Lorentzians. For the fixed-frequency map, 
the fit gave $f_r$ of $\mbox{$906.17 \pm 0.02$ kHz}$ and $w$ of 
$\mbox{$3.61 \pm 0.04$ kHz}$~FWHM with $\chi^2/(N-3)=5$. 
For the frequency-sweep map, the fit gave $f_r$ of 
$\mbox{$906.3 \pm 0.2$ kHz}$ and $w$ of $\mbox{$3.5 \pm 0.2$ kHz}$~FWHM 
with $\chi^2/(N-3)=7$. The errors in $f_r$ and $w$ of the 
frequency-sweep map are again dominated by the frequency sweep's size. 
Note that, for \textbf{\emph{bunched}} beam, both the $f_r$ and $w$ 
results from the fixed-frequency map and the frequency-sweep map 
are consistent. Averaging the fixed-frequency and frequency-sweep 
results gave $\mbox{$f_r = 906.17 \pm 0.02$ kHz}$ and 
$\mbox{$w = 3.61 \pm 0.04$ kHz}$~FWHM  for \textbf{\emph{bunched}} beam. 

Comparing Figs.~\ref{fig:fig1} and \ref{fig:fig2} shows that 
\textbf{\emph{bunching}} the beam increased the resonance's 
width by more than a factor of 2. Moreover, 
both \textbf{\emph{bunched}} maps show full depolarization 
over a wide frequency region around the resonance. Also note that, 
unlike the \textbf{\emph{unbunched}} maps, the two \textbf{\emph{bunched}} 
maps are consistent with each other in both shape and magnitude. 
Finally note that the observed \textbf{\emph{widening}} of the 
\textbf{\emph{proton}} resonance due to bunching is exactly opposite 
to the earlier observed~\cite{narrowres} \textbf{\emph{narrowing}} of 
a \textbf{\emph{deuteron}} resonance due to bunching.

In summary, we recently used an rf solenoid to study the widths of rf spin 
resonances with both \textbf{\emph{unbunched}} and \textbf{\emph{bunched}} 
beams of 2.1~GeV/$c$ polarized protons stored in the COSY synchrotron. 
We first ran the rf solenoid at different fixed frequencies 
near the resonance with the \textbf{\emph{unbunched}} beam. 
We found only a very shallow possible depolarization dip; 
this shallowness may be due to the proton beam's large momentum spread. 
We next made a frequency-sweep map using 400~Hz sweeps centered 
at different frequencies near the resonance; this greatly 
enhanced the \textbf{\emph{unbunched}} dip. 
Then we used the same technique to obtain both fixed-frequency and 
frequency-sweep resonance maps with a \textbf{\emph{bunched}} proton beam; 
these \textbf{\emph{bunched}} maps were consistent 
with each other in both shape and magnitude. We found that the \textbf{\emph{bunched}} 
maps were more than twice as wide as the \textbf{\emph{unbunched}} map. 
Moreover, the \textbf{\emph{bunched}} maps showed full depolarization 
over a wide frequency range near the resonance. 
This \textbf{\emph{proton}} resonance \textbf{\emph{widening}} due to bunching is 
exactly opposite to the recently observed \textbf{\emph{deuteron}} resonance 
\textbf{\emph{narrowing}}~\cite{narrowres} due to bunching.

We thank COSY's staff for a successful run. 
We thank $\mbox{E.D.~Courant}$, $\mbox{Ya.S.~Derbenev}$, $\mbox{D.~Eversheim}$, 
$\mbox{A.~Garishvili}$, $\mbox{R.~Gebel}$, $\mbox{F.~Hinterberger}$, $\mbox{A.~Lehrach}$, 
$\mbox{B.~Lorentz}$, $\mbox{R.~Maier}$, $\mbox{Yu.F.~Orlov}$, $\mbox{D.~Prasuhn}$, 
$\mbox{H.~Rohdje\ss}$, $\mbox{T.~Roser}$, $\mbox{H.~Sato}$, $\mbox{A.~Schnase}$, 
$\mbox{W.~Scobel}$, $\mbox{E.J.~Stephenson}$, $\mbox{H.~Stockhorst}$, $\mbox{K.~Ulbrich}$, 
$\mbox{D.~Welsch}$ and $\mbox{K.~Yonehara}$ for help and advice. The work was supported  
by grants from the German BMBF Science Ministry and its JCHP-FFE program at COSY.

\end{document}